\documentclass[aps,twocolumn,superscriptaddress,prl,floatfix]{revtex4}
\usepackage{amssymb}
\usepackage{amsmath}
\usepackage{graphicx}
\usepackage{xfrac}
\usepackage{color}

\begin{document}
\title{Dynamical Josephson Effects in NbSe$_2$}
\author{S. Tran}
\affiliation{Department of Physics, Joint Quantum Institute and the Center for Nanophysics and Advanced Materials, University of Maryland, College Park, MD 20742, USA}
\author{J. Sell}
\affiliation{Department of Physics, Joint Quantum Institute and the Center for Nanophysics and Advanced Materials, University of Maryland, College Park, MD 20742, USA}
\author{J. R. Williams}
\affiliation{Department of Physics, Joint Quantum Institute and the Center for Nanophysics and Advanced Materials, University of Maryland, College Park, MD 20742, USA}

\date{\today}

\begin{abstract}
The study of superconducting materials that also possess nontrivial correlations or interactions remains an active frontier of condensed matter physics.  NbSe$_2$ belongs to this class of superconductors and recent research has focused on the two-dimensional properties of this layered material. Here an investigation of the superconducting-to-normal-state transition in NbSe$_2$ is detailed, and found to be driven by dynamically-created vortices. Under the application of RF radiation, these vortices allow for two novel Josephson effects to be observed. The first is a coupling between Josephson currents and charge density waves in phase-slip junctions. The second is the Josephson detection of multi-band superconductivity, which is revealed in an anomalous magnetic field and RF frequency response of the AC Josephson effect.  Our results shed light on the nature of superconductivity in this material, unearthing exotic phenomena by exploiting nonequilibrium superconducting effects in atomically-thin materials. 
\end{abstract}
\maketitle

The BSC description of a superconductor provides a mean-field description of Copper pairs in a material in equilibrium. Away from equilibrium novel superconducting phenomena can occur and materials in reduced dimensions can be even more susceptible to nonequilibrium effects~\cite{Tinkham96, Tidecks90}. The advent of atomically-thin materials has ushered new platforms for studying condensed-matter physics in two-dimensions. Recently, there has been a strong thrust towards isolating graphene-like material with interactions between the carriers making possible correlated states at low temperatures.  One such material, NbSe$_2$~\cite{Koloboc16}, has a well-studied past due to the interesting states arising from the competition with a coexisting charge-density wave~\cite{Ugeda16, Lian18} and possible multi-band superconductivity~\cite{Yokoya01, Boakin03, Dvir18}. More recently, NbSe$_2$ has been studied in its two dimensional form, where novel types of superconductivity (Ising)~\cite{Xi16} have been observed. In addition, the presence of spin-orbit interaction, enhanced in two-dimensional due to a broken in-plane mirror symmetry, provides routes to topological phases of matter in this material~\cite{He18, Shaffer19}. 

\begin{figure}[b]
\center \label{fig1}
\includegraphics[width=3 in]{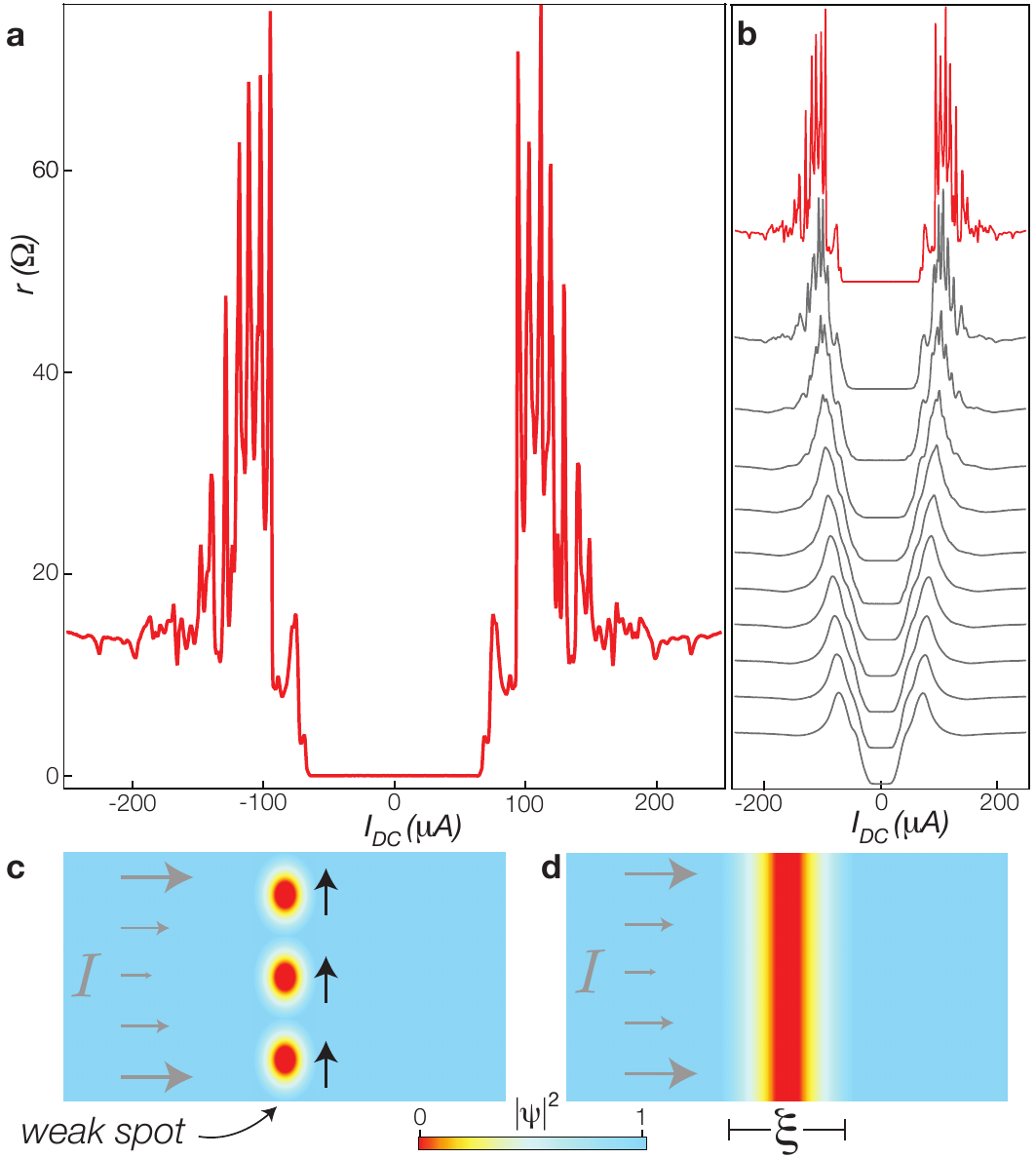}
\caption{\footnotesize{\textbf{a.} Differential resistance $r$ as a function of DC current $I_{DC}$ taken at $B=0$ and $T=1.2\,K$. Rather than a simple superconducting to normal state transition, a series of peaks in $r$ are observed. \textbf{b.} $r$ as a function of $B$ taken between 0 and 1\,$T$ in 0.1\,$T$ steps. The peaks in $r$ die off around $B=0.2\,T$. \textbf{c.} Schematic of the vortex production at the sample edge. Vortices nucleate at points of weak superconductivity (amplitude given in terms of $|\psi|^2$) and propagate perpendicular to the applied current. \textbf{d.} As the production rate of vortices increases, vortices begin to travel in the wake of the preceding, creating a line of suppressed superconductivity.  The color scale in \textbf{c,d} goes from 0 (red, full suppression of superconductivity) to 1 (blue, no suppression).}}
\end{figure}

In this work we fabricate mesoscale devices formed from NbSe$_2$ thin films, focusing on the transition from the superconducting state to normal state as a function of applied DC current ($I_{DC}$), magnetic field $B$ perpendicular to the surface of the NbSe$_2$ and radio frequency (RF) power $P$. The current-driven resistive state is found to be induce dynamic vortices, created at superconducting weak spots under the application of current to the sample. These vortices produce dynamical Josephson effects, which enable the elucidation of two novel properties of the superconducting state in NbSe$_2$. The first is a coupling of the Josephson current and the collective motion of the charge-density wave, resulting in a modification of the Shapiro step diagram observed under the application of microwave radiation. The second is a complete modification of the AC Josephson behavior under RF radiation and magnetic fields.  We discuss possible origins of the latter effect, suggesting it lies in the multiband nature of superconductivity in NbSe$_2$~\cite{Tanaka02}. Each result expands the knowledge of the superconducting state of NbSe$_2$, sheds light on the nature of Josephson junctions in which the superconductivity both competes with other collective phenomena and is multiband in nature. Further, this work opens new paradigms for electronics~\cite{Tanaka10, Tanaka15} and quantum computation applications~\cite{Tanaka14, Tanaka18} in transition metal dichalcogenides. 

In this Article, the superconducting transport properties of four devices (Samples I-IV) of varying thickness are investigated (see \emph{Materials and Methods} and Ref.~\cite{Supp} for details of each device and measurement). In particular, we find that magnetic fields or applied DC currents can drive the system out of equilibrium and these effects can alter the behavior of the superconductor, even in simple measurements like the $I-V$ characteristics. For example, shown in Fig. 1a is the measured differential resistance $r=dV/dI$ versus $I_{DC}$ for a NbSe$_2$ device (Sample I).  As is evident, the pattern is much more complex than a simple transition from a superconducting state to a normal one seen in bulk superconductors. This indicates that another mechanism may drive this transition and modify the transport in the normal state. Prominent in the measured $r$ of Fig 1a is a semi-regular series of peaks. Upon application of $B$, the prominence of these peaks fades at $B$$\sim$200\,mT. The magnitude of this field corresponding to the field at which a regular lattice of flux takes shape~\cite{Maldo13}, suggesting that vortices play a role in the complex pattern seen in $r$ at $B=0$. 

Dynamic incorporation of vortices via the applied current is known to modify the transport in superconductors. In two-dimensional superconductors where the width of the film is larger than both the coherence length $\xi$ and the penetration depth, vortices can play an important role in the transition from a superconductor to a metal~\cite{Tinkham96}. As elucidated in Ref. XX, vortices created near the edge of the sample where the superconducting order parameter is weakest, driving the sample normal. This occurs in two stages XX. First, vortices begin to nucleate at a spot where superconductivity is suppressed and experience a force that is perpendicular to the applied current (Fig. 1c). The vortex travels across the sample and disappears at the other edge or annihilates with an antivortex.   As current is increased, vortex production rate also increases and vortices begin to travel in the wake of the vortex that came before it. This produces a chain of fast (kinematic) vortices, resulting in a region of suppressed superconductivity that spans the sample width (Fig. 1d). This is called a phase slip line (PSL) and it forms a  Josephson junction with a length of $\sim \xi$. The effects of PSLs on DC electrical transport have recently been observed in NbSe$_2$~\cite{Paradiso18}.

\subsubsection{AC Josephson Effect in Phase Slip Lines}

\begin{figure}[b]
\center \label{fig1}
\includegraphics[width=3.5 in]{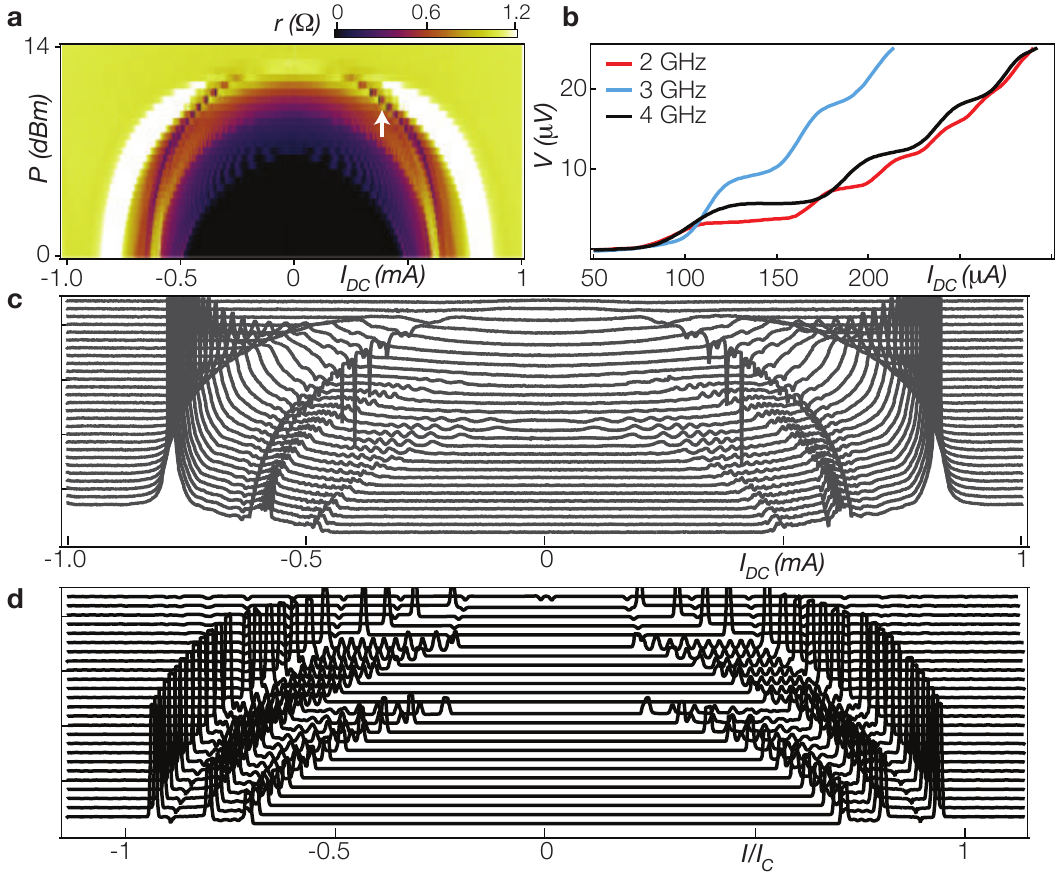}
\caption{\footnotesize{\textbf{a.} Shapiro step diagram $r(I_{DC}, P)$ taken at T=1.2\,$K$. At the appearance of the first PSL, oscillations in $r$ are observed. In addition, a depression of $r$ (indicated with a white arrow where it crosses the transition to the normal state) possess a $P$ dependence different than the PSL. \textbf{b.} The frequency dependence of the step height increases with frequency with a magnitude of $~\sim \sfrac{hf}{2e}$. \textbf{c.} Slices of $r$ at constant $P$ from 2\textbf{a} showing oscillations resulting from the applied RF radiation, matching well with simulations in \textbf{d}. }}
\end{figure}

\begin{figure*}[t!]
\center \label{fig4}
\includegraphics[width=6.75 in]{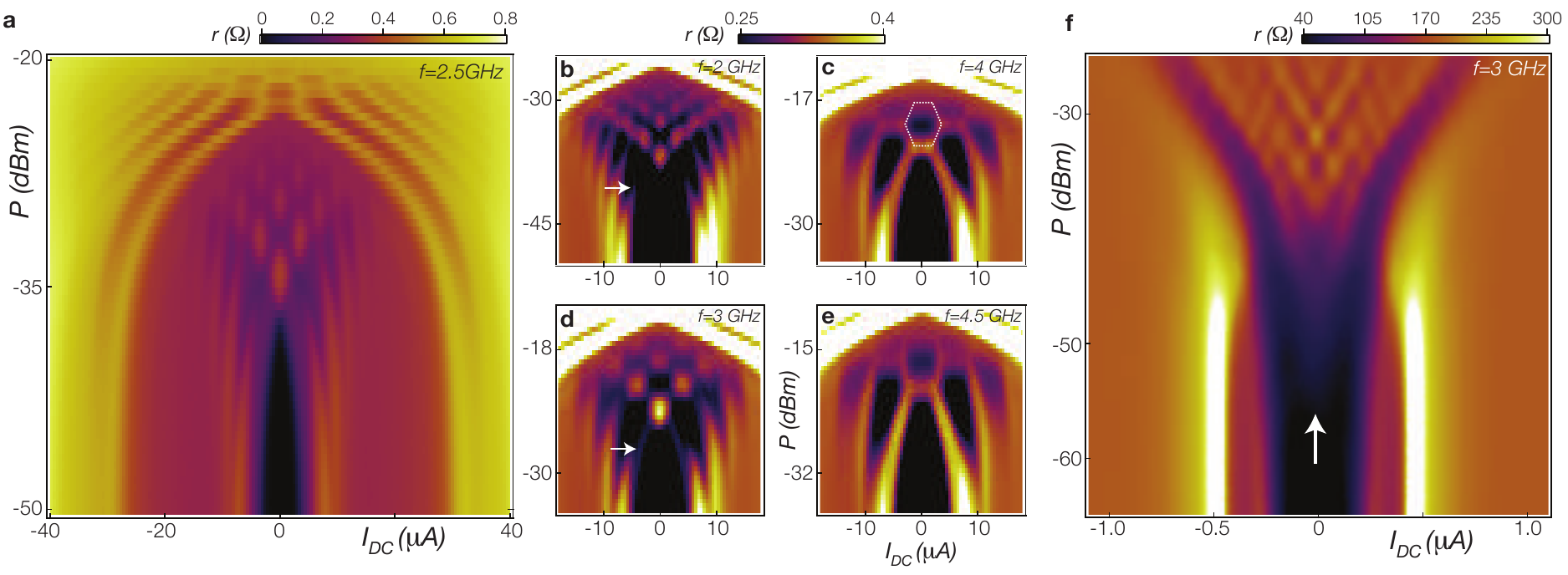}
\caption{\footnotesize{\textbf{a.} Shapiro step diagram for Sample III showing ``inner" and ``outer" Josephson features. \textbf{b-e.} Frequency dependence of the inner features in Sample III showing a Bessel-function-like dependence of the step features. Some unconventional features are observed: a step height which is not independent of power (white arrow) and hexagonal step width (white dashed line). All graphs in b-e have axes $P$ vs. $I_{DC}$. \textbf{f.} Unlike the conventional Josephson effect where the steps originate from critical current, the step features begin at $I_{DC}=0$ (white arrow) at f=3\,GHz in Sample IV.}}
\end{figure*}

We demonstrate superconducting quantum interference across a PSL that creates a Josephson junction by addressing the junction with RF radiation. Should there be a junction created by a PSL, RF radiation mixed with the internal Josephson frequency should produce steps in the $I-V$ curve (or peaks in $r$), or Shapiro steps~\cite{Barone}. Fig 2a show $r$ for Sample II as a function of the $I_{DC}$ and $P$, at a frequency $f$=3\,GHz. RF radiation is applied at the top of the cryostat to a lead of our sample (i.e. it does not include the attenuation incurred from the electrical path down to the sample). Plots of this type are referred to as Shapiro step diagrams in this Article. A regular pattern of peaks in $r$ are observed. The area under the peaks in $r$ (corresponding to the step height in a I-V curve) have a frequency dependence. As shown in Fig. 2b, the step height of the $I-V$ curve obtained from integration of $r$ produce steps heights of approximately $2\,\mu$V/GHz, identifying the peaks as Shapiro steps arising from the junction formed by the kinematic vortices. 

Two key features help to explain the deviations from a Shapiro step diagram common to conventional Josephson junctions. The first is that the Josephson coupling only turns on at a finite current associated with the formation of a PSL. The data of Fig. 2a allows for the identification of two PSLs corresponding to the transition from black to purple (at $|I_{DC}|$=500\,$\mu$A at $P=0$\,dBm) and from purple to yellow (at $|I_{DC}|$=600\,$\mu$A  at the same power).  It is only after these slips form that oscillations in $r$ are observed. A broad white feature around $|I_{DC}|$=800\,$\mu$A marks the transition to the normal state. One further feature, a dark depression of $r$, is distinct in the the data of Fig. 2a in that the power dependence follow a different trajectory than the power dependence of the PSLs and the transition to the normal state (this feature is indicated by a white arrow on the positive current side). This feature is not captured in the simulation of the Shapiro diagram using just PSLs. 

In the normal state, NbSe$_2$ undergoes a charge density wave transition at a temperature of 33\,K in the bulk and a temperature that is layer-number dependent in thin films~\cite{Xi15}. Motion of a sliding CDW is described by an equation that is very similar to Josephson junctions. For a Josephson junction weak-link comprised of a CDW material, coupling between the the Josephson current and the sliding motion of the CDW can affect the $I-V$ curves when the length of the junction (the distance between the two superconducting leads) is of order a coherence length~\cite{Visscher97}. For NbSe$_2$, this coherence length is $\sim$10\,nm~\cite{Bana13}, a length scale difficult to achieve by conventional lithographic techniques. However, here the junction length is automatically set to this length scale, making the results of Ref.~\cite{Visscher97} applicable to our junctions. Simulations derived from the governing equations of a CDW weak link Josephson junction as: 
\begin{equation}
	\label{Current}
	I = I_C \sin \varphi + \frac{e N}{\pi} \dot{\chi} + \frac{\hbar}{2eR_{N}} \dot{\varphi} + I_{AC} \sin \Omega t,
\end{equation}

\begin{equation}
	\label{voltage}
	\frac{\hbar \dot{\varphi}}{2 e} =V_T \sin \chi + \frac{e}{\pi} N R_c \dot{\chi},
\end{equation}

\noindent where $I_C$ and $\varphi$ are the critical current and phase difference across the junction, $I_{AC}$ and $\Omega$ are the AC drive amplitude and frequency, $V_T$ is the threshold voltage for CDW motion, $R_C$ is the dissipation the CDW in motion, $N$ is the number of 1D CDW chains, and $\chi$ is the phase of the CDW (see Ref.~\cite{Supp} for more detail). Using the values of $R_C$=20, $V_T$ = 0.8, $N$ = 1, $I_C = R_N$ = 1, $\hbar$ = 1, $e = 1$, $\Omega$ = 0.03, simulations show a dip in $r$ between the second phase slip line and the transition to the normal state that has a power dependence distinct from the superconducting features of the diagram. Hence we correlate this dip with CDW existing in the normal region of the Josephson junction and a mode-locking between the CDW and the Josephson current (i.e. between $\varphi$ and $\chi$). 

\subsubsection{Anomalous AC Josephson Effect Induced By Vortex Motion}

Moving to the study of two thinner samples (Sample III and IV), more structure is generated for a device subjected to DC currents and RF radiation. Observed in Fig. 3a for Sample III are two distinct collections of oscillations of $r$. The first occurring at the transition identified as the formation of a PSL when the device is measured in the absence of RF radiation. This transition is visible in Fig. 3a at a value of $|I_{DC}|$=30\,$\mu A$ at P=-45dBm, following a power dependence similar to what was observed for the PSL in Fig. 2a. In addition, there are features at smaller values of $| I_{DC} |$.  As will be seen below, these features which occur at lower values have a quantitative different Josephson behavior than the ones at larger values of  $| I_{DC} |$. Henceforth, we refer to the unconventional Josephson features as ``inner" features (as they occur at smaller values of $| I_{DC} |$) and those with a conventional Josephson effects as ``outer" features. 

To establish that these peaks in $r$ originate from a Josephson-like relation, the frequency dependence of the peaks are shown in Fig. 3b-e. As the frequency increases, the peaks become more pronounced and the $I_{DC}$ spacing between peaks increases. The pattern is reminiscent of the Bessel function dependence of the Shapiro step width that dictate the pattern observed in conventional Shapiro diagrams~\cite{Barone} The power and frequency dependence of the shape of this pattern is set by a characteristic frequency $f_c=2eV_C/h=2e I_CR_N/h$.  Hence, it is possible to extract the approximate strength of the superconducting order parameter associated with the Josephson-like phenomena without an explicit value for either $I_C$ or $R_N$ for the inner features. By matching the frequency dependence of the patterns, we are able to extract an $I_CR_N$ product of 14$\mu V$ for Sample III and 44$\mu V$ for Sample IV. For conventional Josephson junctions, this product is a proxy for the strength of the coupling between the superconductors forming the Josephson junction.

\begin{figure}[t!]
\center \label{fig4}
\includegraphics[width=2.75 in]{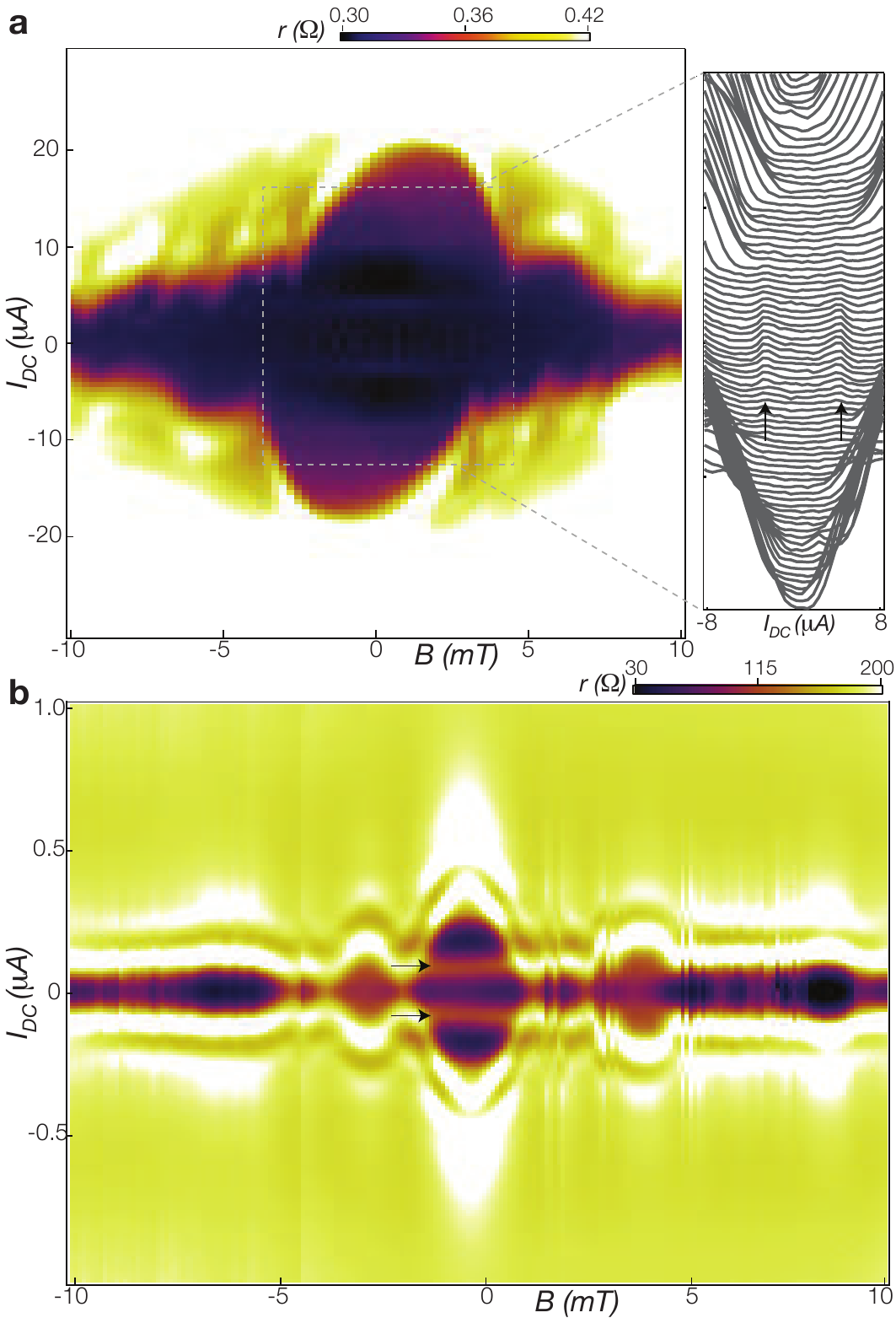}
\caption{\footnotesize{\textbf{a,b.} Magnetic field dependence of the Shapiro steps for Samples III ($f$=5\,GHz, $P$=-17.5\,dBm) and IV ($f$=7\,GHz, $P$=-10\,dBm). In addition to the Fraunhofer-like pattern observed in the outer features, a magnetic field independent peaks in $r$ are observed for the inner features in each sample for values of applied flux less than $\Phi_o$ (indicated in each figure by black arrows). The peaks in \textbf{a} are faint, so cuts of constant $B$ the figure are shown on the right. }}
\end{figure}

The pattern observed is, however, only qualitatively similar to conventional junctions. First, the integrated peak height is $\sim$10 times smaller than is expected. Another prevalent deviation is the hexagonal shape in the Shapiro steps, most prominently seen in the 4 (Fig. 3d) and 4.5 GHz (Fig. 3e) data. Further, there is also an unconventional power dependence of the peak width, essential in extracting the peak height from measurements of the differential resistance $r$. Unlike conventional junctions where the peak height is uniform, here we observe regions where the peak height almost disappears, seen most clearly in Fig. 3b,d (indicated by the white arrow). This makes an exact extraction of a precise peak height difficult. The peak heights do, however, increase with increasing frequency as expected for Josephson junctions. Finally, the evolution of the step features as a function of $P$ proceeds in an unexpected manner, as seen in Sample IV. Rather than peeling off from the superconducting to normal state transition, the inner features at $f$=3\,GHz originate from a point at $I_{DC}=0$ (white arrow, Fig. 3f) . For comparison, plots of the Shapiro step diagram for conventional Josephson junctions at low and high values of $f_c$ are given in the Supplemental Material~\cite{Supp}.

\begin{figure*}[t!]
\center \label{fig5}
\includegraphics[width=6.75 in]{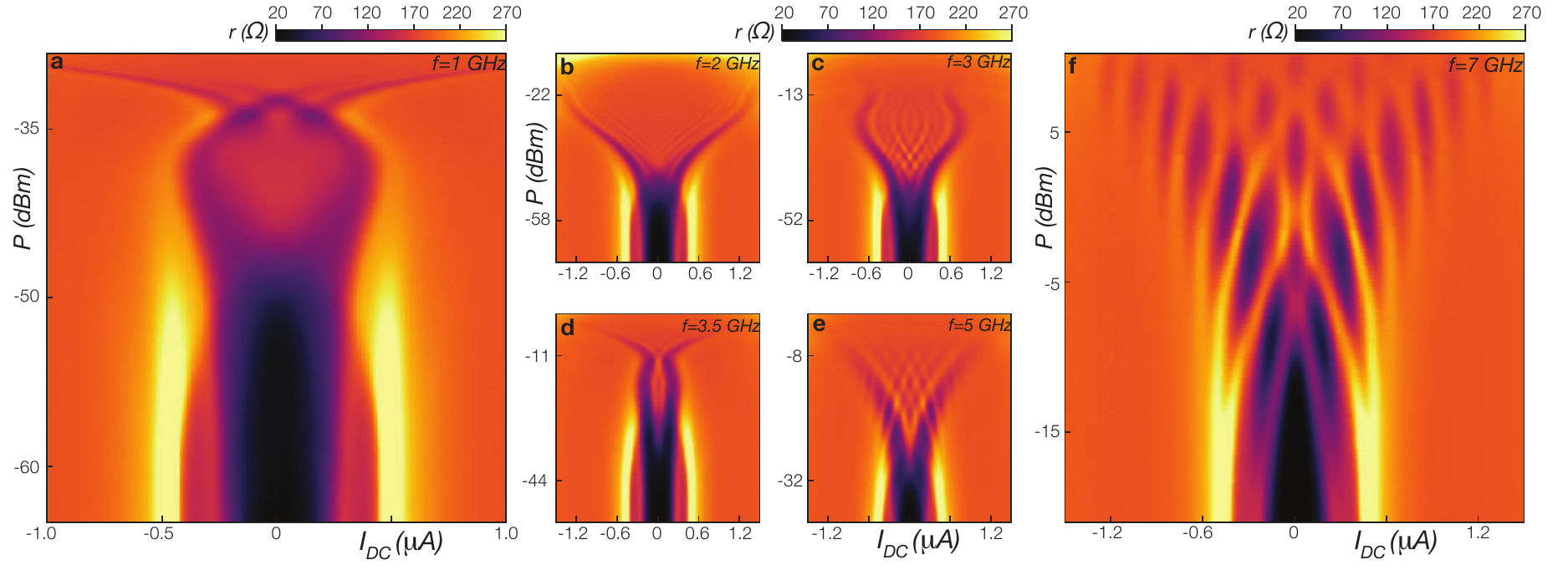}
\caption{\footnotesize{\textbf{a-f.} Shapiro step diagrams for Sample IV demonstrating the frequency dependence of Josephson features for $f$ from 1\,GHz to 7\,GHz. All plots have axes $P$ vs. $I_{DC}$, with a. being plotted on a smaller frequency range than the b-e. The Shapiro steps are shown to shrink and then expand with increasing $f$, contrary to the conventional Josephson effect where Shapiro steps only get spread out with increasing $f$. The Shapiro step diagrams also exhibit different structures when comparing a-c ($f < 3.5$\,GHz) to e-f ($f > 3.5$\,GHz).}}
\end{figure*}

Perhaps the most striking distinction between the inner and outer features is their respective dependence on an applied magnetic field. Contrary to the conventional Josephson effect, the inner features demonstrate an insensitivity to an applied magnetic field.  By contrast, the outer features exhibit a conventional magnetic diffraction pattern expected for long Josephson junctions. This contrasting behavior in magnetic field is displayed in Fig. 4a,b for Samples III and IV. The plots are taken under RF radiation of 5\,GHz (Fig. 4a) and 7\,GHz (Fig. 4b). The outer features associated with the PSL show a diffraction pattern expected for Josephson junctions (Fig. 4a,b). The inner features however, display an entirely different behavior. Clear field-independent peaks in $r$ (indicated by black arrows) that correspond to the RF generated peaks are observed over the range of magnetic field of one quantum of flux, measured by the first node in the magnetic diffraction pattern of the outer features. Beyond this field, it is difficult to differentiate the inner and outer features. 

A third departure from the conventional Josephson effect is the the frequency dependence of the Shapiro step diagram. The frequency dependence of the Shapiro steps are summarized in Fig. 5 for $f$ ranging from 1\,GHz to 7\,GHz (see Ref.~\cite{Supp} the entire collection of data sets in this frequency window and to see the frequency-dependent Shapiro step diagrams for a conventional Josephson junction) for Sample IV. What is observed is that the ``envelope" outlining these features display a nonmonotonic dependence on $P$, most prominently observed for $f$=1 (Fig. 5a) , 3 (Fig. 5b), and 3.5\,GHz (Fig. 5c): the features first expand, then shrink towards smaller values of $I_{DC}$ and then spread outwards again. This is contrasted by the more conventional envelope shape as a function of $P$ seen in Fig. 5e,f, where the envelope size continues to increase as $P$ is increased. It is also of note that only in the vicinity of $f$=3\,GHz does the step features originate at $I_{DC}$=0: for other frequencies the step features can be seen to originate in at the normal state to superconducting transition, as is observed in conventional junctions. 

We now discuss possible origins for this new type of Josephson effect observed. Two possible candidates are first identified for the origin of this effect: multiple phase slips (Josephson junctions) induced by disorder in NbSe$_2$ and Shapiro steps originating from the CDW. We rule out the first these possibilities. An additional PSL is indeed possible, however, cannot explain the observed behavior. For example, the a phase-slip Josephson junction has a complex dependence on magnetic field and does not have the frequency dependence observed~\cite{Siva03}. Josephson behavior has been observed in CDW systems: the similarity between the equations governing the CDW and Josephson junction dynamics allows for Shapiro steps to be observed in materials like NbSe$_3$~\cite{Zettl84}. The primary difference is that the dynamical equation for CDW is one for the voltage produced by the sliding mode of the CDW. Hence, although the pattern is also described by a Bessel function dependence on RF power, the differential resistance is locally a maximum in the regions of the Shapiro steps, separated by regions of lower resistance~\cite{Supp}. Further, there is no anomalous frequency dependence as we observed. We also note that the AC-dependent voltage step height is of order 10\,mV/GHz~\cite{Zettl84}, roughly three orders of magnitude higher than expected for Josephson junctions and four orders of magnitude greater than the typical peak height is observed in Sample III and IV. 

The magnetic-field independence and anomalous frequency dependence suggests that dynamical mode locking distinct from that in a Josephson junction is responsible for the observation. Another mechanism for the production of Shapiro steps is possible in superconductors: the collective motion of a vortex lattice in a periodic potential also gives rise to Shapiro steps~\cite{Fiory71, Look99}. Hence we turn towards the production of vortices at the edge of the sample (Fig. 1c) as a source of the steps. The dynamics of current-induced vortices would not be dependent on the small values of $B$ explored in Fig. 4, therefore would explain the field independence observed in experiment. Further, the dynamics of vortex motion in a periodic potential is often described by the Frenkel-Kontrova model. Numerical simulations of this model has demonstrated oscillations of the step width and depinning current (the envelope described above) as a function of frequency~\cite{Hu07}. Specifically, oscillations in these two features occurs in the low-frequency regime ($f <0.16$, becoming monotonic as frequency is increased. This is consistent with the experimental observations.  Sample III ($f_C$=6.7GHz) is explored in the frequency regime above the oscillation region ($f$ relative to the characteristic frequency is explored in the range 0.18 to 0.75). Sample IV ($f_C$=21.2GHz, explored in the range 0.05 to 0.28) observe the crossover expected from theory: oscillations are observed in the envelope until $f$=4\,GHz (0.18 relative to the characteristic frequency) after which the oscillations exhibit a monotonic dependence on frequency is measured. 

Therefore, using the independence vortex of production to small magnetic fields and Frenkel-Kontrova model accounts well for the anomalous behavior observed in Samples III and IV. However, clarification is needed since it is not immediately clear where the periodic potential is for the vortices traversing the sample. Further, step heights measured for coherent vortex motion in a periodic potential must be equal to or larger than what is expected for Josephson junctions. Measurements here observe a smaller value of step height. 

The dynamics of sample-edge vortex production is markedly different in multiband superconducting materials. For a superconductor with two-bands, a vortex is a composite of two vortices, one from each band~\cite{Tanaka02}. The widing numbers on each band can be different, allowing for fractional flux to be associated with each vortex~\cite{Tanaka02, Babaev02}. However, an attractive force between vortices from different bands pins the two vortices together in a composite vortex with a single quantum of flux when the superconductor is in equilibrium~\cite{Lin13}. When the superconductor is driven from equilibrium, dissociation of the composite vortex is possible. 	

In our samples, dissociation can happen for two reasons. First the Bean-Livingston barrier for entry into the superconductor is, in general, different for the two bands, hence can promote dissociation at the sample edge~\cite{Silaev11}. Second, once in the sample, the velocity of vortex propagation will be different for each vortex associated with the two bands~\cite{Lin13}. Thus, even if a composite vortex enters the sample, dissociation can occur do to different propagation velocity. In the latter, Shapiro steps from the Frenkel-Kontrova model have been predicted: the faster fractional vortices will pass over the periodic potential provided by the slower vortices. The step height in this scenario is set by the difference in propagation velocity, rather than the actual velocity of the vortex. This allows for the reduction of the step height in proportion to (1-$v_2/v_1$)~\cite{Lin13}. The relative velocity of the two bands is given by $v_2/v_1=(\xi_1/\xi_2)^2=(\Delta_2/\Delta_1)^2$. The ratio of the superconducting order parameter of the two Nb-derived bands has been measured by ARPES, finding a ration of 0.9 between these two bands~\cite{Yokoya01}. The would produce an 80\% reduction of the Shapiro step height, consistent with our observations. 

\subsubsection{Conclusions}
In the communication, we have detailed the transition from the superconducting to the normal state. We have found this transition is driven by vortices. This allowed for the elucidation of two phenomena which previously have not been observed. The first is a coupling of the supercurrent in a Josephson junction to the sliding motion of a charge density wave. This was facilitated by the small length of a phase-slip Josephson junction. The second, observed in thinner samples, was the interplay between vortex production at the edge of the sample, giving rise to Shapiro steps that were smaller than expected, insensitive to magnetic field and possessing an anomalous behavior as a function of frequency. At the heart of this observation was the dissociation of composite vortices in a multiband superconductor driven from its equilibrium state. Both of these results furthers the understanding of NbSe$_2$ and demonstrates the power of investigating superconducting materials out of equilibrium. \\

\emph{Acknowledgements:} We thank Sergiy Krylyuk and Albert Davydov for technical discussion including materials preparation. This work was sponsored by the National Science Foundation Physics Frontier Center at the Joint Quantum Institute (PHY-1430094).

\emph{Author contributions:} S. T. and J. S. fabricated the samples and performed the measurements. All authors discussed the data and contributed to the manuscript.

\emph{Methods:} The 2H-NbSe$_2$ crystals used in this study were commercially obtained from HQ Graphene. The bulk crystals have $>$ 99.995\% purity and were characterized using XRD, Raman, and EDX. We mechanically exfoliated thin flakes from the bulk crystals on Si/SiO$_2$ substrates using scotch tape. Flakes of suitable geometry and thickness were identified using optical microscopy. Once flakes were identified, polymethyl-methacrylate (PMMA) is spin-coated onto the surface of the Si/SiO$_2$ substrate in preparation for patterning of leads. Our devices consisted of four leads with a micron spacing that were patterned using electron-beam lithography. Since we do not encapsulate our flakes, we limit their exposure to air by performing the fabrication and device cooldown in less than 5 hours after exfoliation.

All measurements were performed in a cryostat capable of fields to $B$=10\,T. All lines were equipped with electronic filtering in the range 10\,kHz to 10\,GHz except for the high frequency coax line used to supply the sample with the RF radiation. Measurements of the current-bias differential resistance were performed using standard lock-in techniques at 11Hz. All measurements were performed at $T$=1.2\,K except for Sample IV which was taken at $T$=50\,mK. RF radiation is applied to a single lead of the device through an on-chip bias tee. 

Additional information can be found in the Supplemental Material~\cite{Supp}

\emph{Data Availability:} The data and simulation code that support the plots within this paper and other findings of this study are available from the corresponding author upon reasonable request.

\end{document}